\documentclass[twocolumn,aps,prl,showpacs,amsmath,amssymb,floatfix]{revtex4}
\usepackage{eucal,bm,graphicx}
\begin{document}
\title{Inverse turbulent cascades and conformally invariant curves}
\author{D. Bernard$^{1}$, G. Boffetta$^{2}$,  A. Celani$^{3}$
and G. Falkovich$^{4}$}
\affiliation{
$^{1}$ LPT-ENS,  24 Rue Lhomond, 75231 Paris Cedex 05, France\\
$^{2}$ Dip. di Fisica Generale and INFN, Univ.
di Torino,
via Pietro Giuria 1, 10125 Torino, Italy\\
$^{3}$ CNRS, INLN, 1361 Route des Lucioles,
06560 Valbonne Sophia Antipolis, France\\
$^{4}$ Physics of Complex Systems, Weizmann Institute of Science,
Rehovot 76100, Israel }
\begin{abstract}
We offer a new example of conformal invariance far from
equilibrium --- the inverse cascade of Surface Quasi-Geostrophic
(SQG) turbulence. We show  that temperature isolines are
statistically equivalent to curves that can be mapped into a
one-dimensional Brownian walk (called Schramm-Loewner Evolution or
SLE$_{\kappa}$). The diffusivity is close to $\kappa=4$, that is
iso-temperature curves belong to the same universality class as
domain walls in the $O(2)$ spin model. Several statistics of
temperature clusters and isolines are measured and shown to be
consistent with the theoretical expectations for such a spin
system at  criticality. We also show that the direct cascade in
two-dimensional Navier-Stokes turbulence is not conformal
invariant. The emerging picture is that conformal invariance may
be expected  for inverse turbulent cascades of strongly
interacting systems.
\end{abstract}
\pacs{47.27.-i}
\maketitle

To identify underlying symmetries is a central problem of the
statistical physics of infinite-dimensional strongly fluctuating
systems. Turbulence is a state of such a system which is deviated
far from equilibrium and is accompanied by dissipation. Excitation
and dissipation usually break symmetries such as scale invariance,
isotropy and time reversibility. In fully developed turbulence,
the scales of excitation and dissipation differ strongly and are
separated by the so-called inertial interval. The main fundamental
problem of turbulence is how universal
is the statistics of fluctuations in the inertial interval
\cite{FS}. Are symmetries, broken by excitation and
dissipation, eventually restored in that range \cite{Pol}?
Cascades can be direct or inverse depending on whether
the integral of motion is
transferred towards small or large scales, respectively. Symmetries
broken by excitation (scale invariance and isotropy) are
generally not restored in direct cascades due to the existence of
statistically conserved quantities 
\cite{FS,FGV}.
On the contrary, symmetries are expected to be restored  in the
inverse cascade where one looks at scales much larger than the
pumping scale.
This is consistent with the observation that inverse cascades are scale
invariant \cite{BCV00,Tab,CCMV}. Moreover, it has been shown
recently that the statistics of zero-vorticity lines in the
inverse cascade of two-dimensional (2d) Navier-Stokes turbulence
display conformal invariance (i.e. local scale invariance),
revealing an unexpected connection with percolation \cite{BBCF06}.
Such tantalizing results, while awaiting a theory capable of
explaining them
 from first principles, pose new questions:
Are there other turbulent flows that share this property and do
they belong to the same universality class of percolation? In this
Letter we answer them by a numerical investigation of SQG
turbulence. This system
 is relevant for
geophysical applications \cite{HPGS95} and qualitatively similar
to  Navier-Stokes 2d turbulence. We show that zero-temperature
isolines are SLE$_4$ curves at large scales (in the inverse
cascade). Therefore, the isolines belong to the same universality
class as the trace of a harmonic explorer, certain isolines of a
Gaussian (free) field \cite{SS}, interfaces in the $O(2)$ model
and frontiers of Fortuin-Kasteleyn clusters in the $4-$state Potts
model at the critical point (for an introduction to SLE and
statistical models see Refs.~\cite{C05,BB06} and references
therein). This connection allows to obtain analytical predictions
for some characteristic exponents of cluster and loop statistics
that compare well with numerical results.

\begin{figure}[!hb]
\includegraphics[angle=270,width=8.5cm]{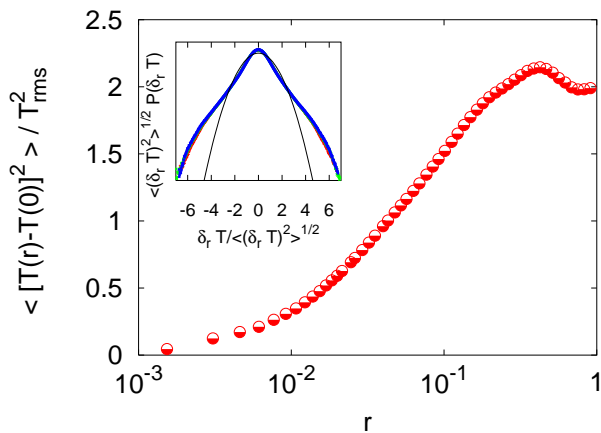}
\caption{The second-order structure function of temperature
differences and, in the inset, probability density functions for
$r=0.02,0.04,0.06$ compared to a Gaussian density (solid line).
Data have been obtained by a direct numerical simulations of
(\ref{eq:1}) by a pseudo-spectral code in a fully periodic, square
domain of size $1$ with $4096^2$ lattice points. Gaussian
white-in-time forcing $f$ has correlation length
$l_f\approx1/200$. The system is kept in a statistically
stationary state by supplementing (\ref{eq:1}) with a linear
damping term $-T/\tau$ that models bottom friction and extracts
energy at very large scales $l_\tau \propto \tau$,  ($l_\tau =1/20
\div 1/10$ depending on  $\tau$). } \label{fig:0}
\end{figure}

The SQG model describes a rotating stably stratified fluid with a
uniform potential vorticity~\cite{HPGS95}. The temperature is
advected along a surface bounding a constant potential vorticity
interior,
\begin{equation}
\partial_t T + {\bm v} \cdot {\bm \nabla} T = \varkappa \nabla^2 T +
f\,, \label{eq:1}
\end{equation}
and determines the velocity ${\bm v}=\hat{\bm
z}\times{\bm \nabla}\psi$, $\psi({\bm x},t)=\int d{\bm y} \,|{\bm x}-{\bm
y}|^{-1} T({\bm y},t)$.
Without dissipation  and forcing ($\varkappa=f=0$) the equations
admit two positive-defined quadratic invariants $Z=\int T^2 d{\bm
x}$ and $E= \int \psi T d{\bm x}$. In the presence of a forcing
that injects temperature fluctuations at a scale $l_f$, a double
cascade develops, akin to the one observed in 2d Navier-Stokes
turbulence: the ``energy'' $E$ flows upscales whereas the
``enstrophy'' $Z$ goes downscales. Here we focus on the inverse
cascade. Requiring the energy flux 
to be scale independent one gets the scaling law
$\delta_r T=T({\bm x}+{\bm r})-T({\bm x}) \sim r^H$ with $H=0$ (i.e.
logarithmic correlation functions). Indeed, numerical simulations show that the
temperature field in the inverse cascade displays a self-similar
statistics with a scaling compatible with dimensional expectations
(see \cite{HPGS95,PHS94,S00,SBHMTHV02,CCMV04} and Fig.~\ref{fig:0}).
\begin{figure}[b]
\includegraphics[width=6cm]{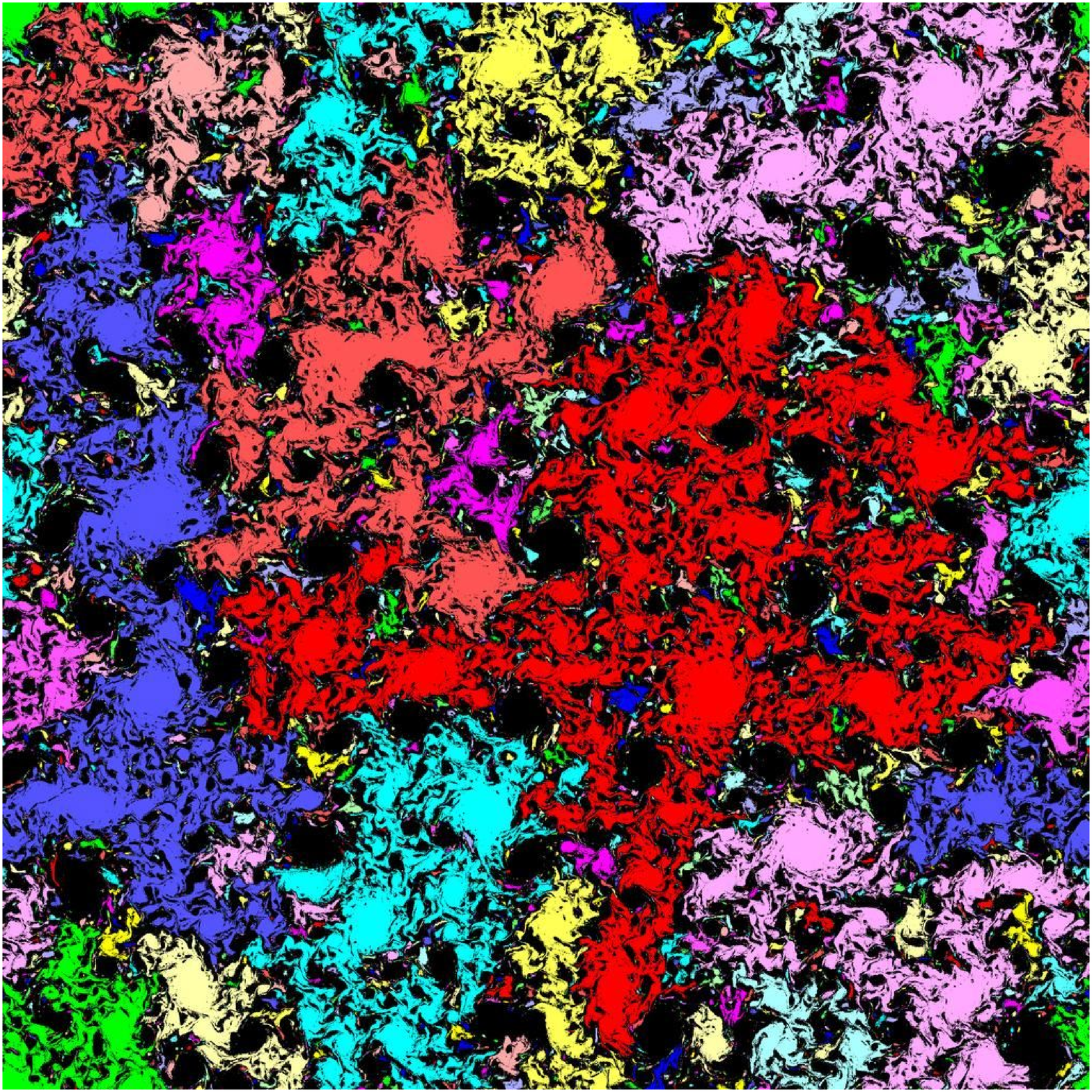}
\caption{Temperature clusters in the inverse cascade of SQG
turbulence. These are  connected domains with positive temperature.
Negative temperature regions are  black. } \label{fig:1}
\end{figure}

\begin{figure}[b]
\includegraphics[width=8.5cm]{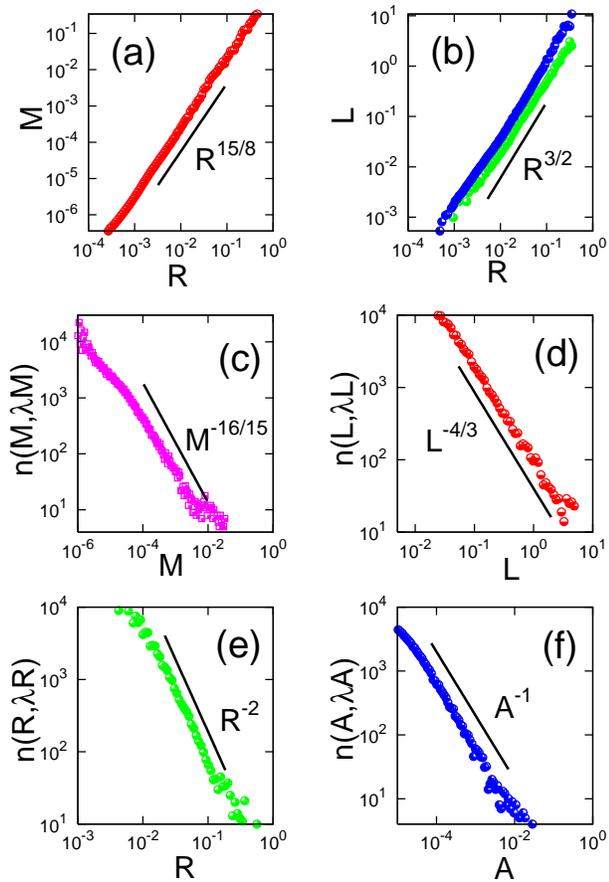}
\caption{Cluster and loop statistics for SQG turbulence. (a) The
average area $M$ versus the radius of gyration $R$. (b) The length
of a loop (blue symbols) and its externally accessible perimeter
which is obtained by subtraction of fjords with necks smaller than
$l_f$ (green symbols) versus $R$. (c) Number of clusters of area
between $M$ and $\lambda M$. (d) Number of loops of length between
$L$ and $\lambda L$. (e) Number of loops of radius between $R$ and
$\lambda R$. (f) Number of loops of area between $A$ and $\lambda
A$. In all figures $\lambda \simeq 1.1$. The solid lines are the
theoretical expectations for the $O(2)$ model.} \label{fig:2}
\end{figure}

We now consider the connected regions of like-sign temperature
(clusters) and their boundaries (loops) --- see Fig.~\ref{fig:1}.
To guess the cluster statistics one needs the knowledge of the
scaling properties of the temperature field. Indeed,  for a
self-similar field with Hurst exponent $H=0$ the fractal dimension
of loops is $3/2$ \cite{KH95}. If one assumes that such loop
ensemble has a conformal invariant scaling limit, it should belong
to the same universality class as loops in the $O(2)$ model in the
dense phase. By exploiting the Coulomb gas representation of the
latter system (with $g=1$, \cite{N84}) and general scaling
arguments \cite{KH95} it is possible to derive analytically a set
of scaling exponents associated to cluster and loop statistics.
These include the fractal dimensions of clusters and loops, the
power-law exponents for the number of clusters of given mass and
the number of loops of given length, radius of gyration, or area.
In Figure~\ref{fig:2} such statistics are displayed for Surface
Quasi-Geostrophic turbulence and shown to be consistent with that
of  $O(2)$ model.

\begin{figure}[b]
\includegraphics[width=6cm]{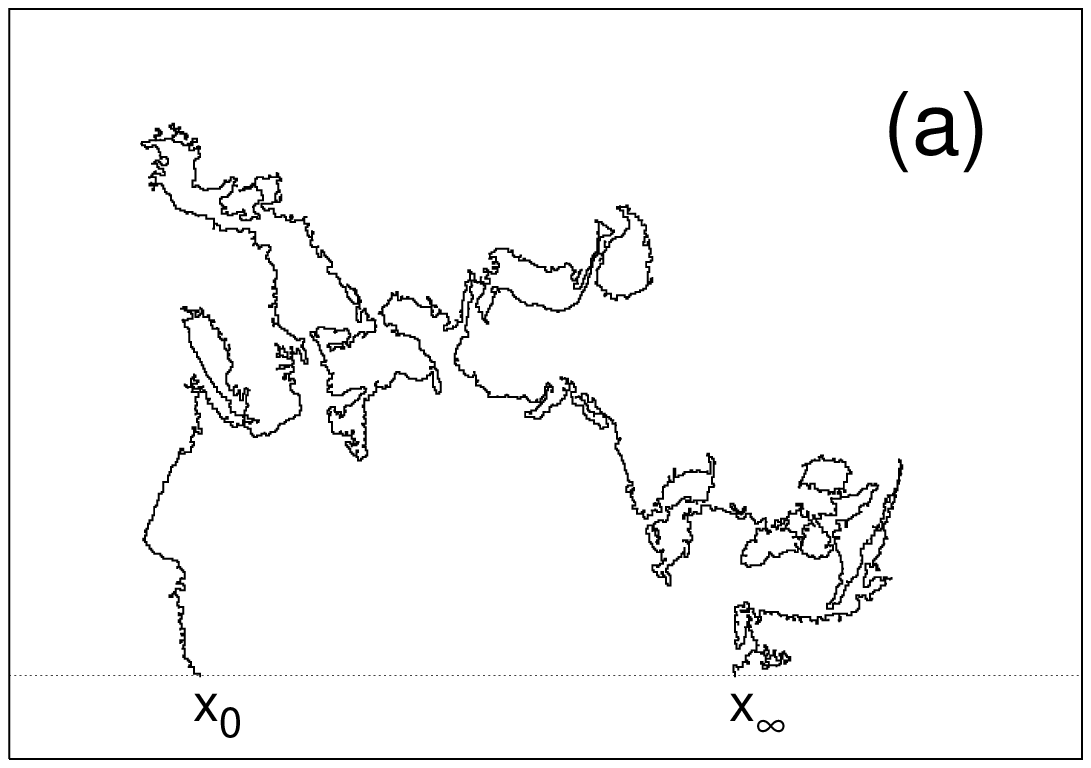}
\vspace*{0.1cm}\\
\includegraphics[width=6cm]{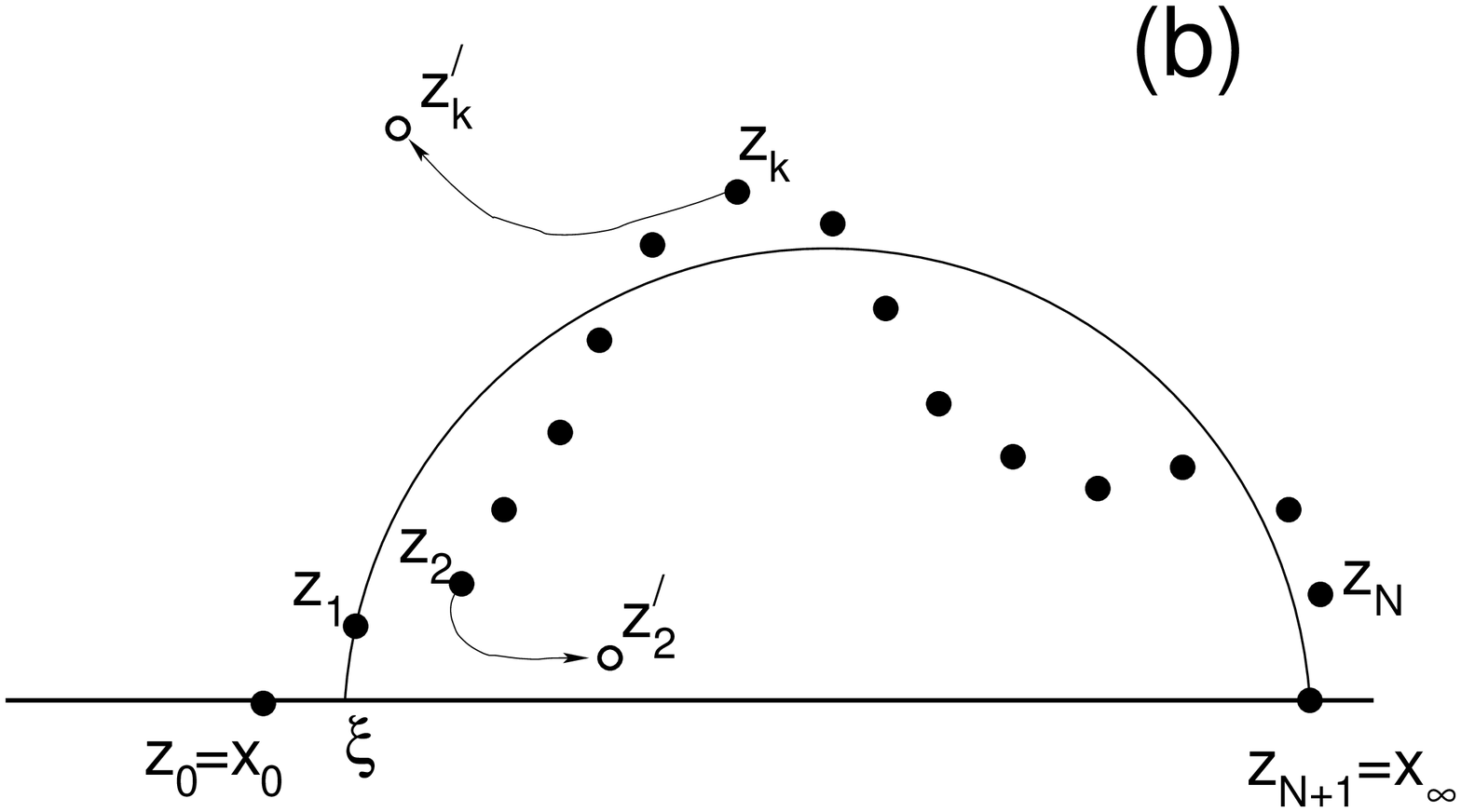}
\caption{(a) A putative SLE trace. (b) A cartoon of the algorithm
used to extract the driving function from the  trace (see text).
The  procedure has been checked on an ensemble of
self-avoiding loops where it yielded the correct value
$\kappa=8/3$ with an uncertainty of less than 5\%.} \label{fig:3}
\end{figure}

These results give a strong indication that loops, i.e.
zero-temperature isolines, might be conformally invariant objects.
If this is the case they should be statistically equivalent to
SLE$_\kappa$ curves with the diffusivity $\kappa=4$, as it is
conjectured for the $O(2)$ model. To verify directly this
hypothesis, we proceeded as follows. First, we  identify putative
SLE traces. After having isolated zero-field lines, a walker follows them
as they explore the upper-half plane.
The search ends when the walker touches the real axis at a distance
from the origin larger than $l_\tau$.
A sample trace is shown in Figure~\ref{fig:3} (a). This selection
procedure,  self-consistent for $\kappa \le 4$, yields a set of
curves in the half-plane, which  are expected in the scaling
limit to converge to chordal SLE joining two points on the real
axis. Second, we  extract the Loewner driving function from the
trace. To this aim, let us consider chordal SLE in the upper-half
plane $H$ from $0$ to $x_\infty$. We parametrize the curve by
the dimensionless $t$, not to be confused with time in
(\ref{eq:1}). The  equation for $g_t(z)$ which maps the half-plane
minus the trace up to  $t$ into $H$ itself, is $\partial_t g_t =
2/\{\varphi^{'}(g_t)[\varphi(g_t)-\xi_t]\} $ where
$\varphi(z)=x_\infty z/(x_\infty-z)$.
The equation for $g_t$ can be solved for a constant $\xi$: $
G_{t,\,\xi}(z)=x_\infty \{\eta x_\infty (x_\infty-z)+
[x_\infty^4(z-\eta)^2+4t(x_\infty-z)^2(x_\infty-\eta)^2]^{1/2}\}/
\{x_\infty^2 (x_\infty-z)+
[x_\infty^4(z-\eta)^2+4t(x_\infty-z)^2(x_\infty-\eta)^2]^{1/2}\}
$ where $\eta=\varphi^{-1}(\xi)$.
In this case the trace is the semicircle joining $\eta$ and
$x_\infty$.
\begin{figure}[!b]
\includegraphics[angle=270,width=8.5cm]{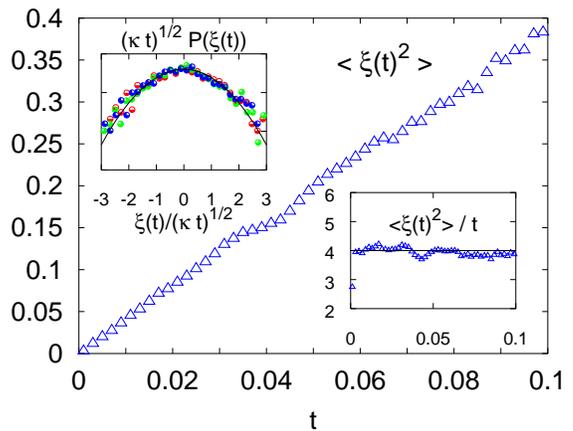}
\caption{Statistics of the driving function. In the main frame is shown
the diffusive behavior of $\xi(t)$. Lower-right inset:
the diffusion coefficient is $\kappa=4 \pm 0.2$. Upper-left inset:
the probability density of $\xi_t$, rescaled by its variance is Gaussian
(different symbols correspond to $t=0.02, 0.04, 0.08$).}
\label{fig:4}
\end{figure}
For a generic $\xi(t)$ we  partition the  interval $[0,T]$
into subintervals $[t_n,t_{n+1})$ with 
$t_0=0$, $t_{N+1}=T$ where we approximate the driving function by
the constant $\xi_n=\xi(t_n)$ and express $g_t$ as a composition
$G_{t_N-t_{N-1},\,\xi_{N-1}} \circ \cdots \circ G_{t_1,\,\xi_0}$. It
is now possible to extract the driving function from a candidate
SLE trace, approximated by a sequence of points
$\{z_0,z_1,\ldots,z_{N+1}\}$, where $z_0=0$ and
$z_{N+1}=x_\infty$. The first step is to identify the unique
semicircle passing through the points $x_\infty$ and $z_1$ [see
Fig.~\ref{fig:3} (b) for an illustration]. This yields the values
for
$\eta_0=\varphi^{-1}(\xi_0) =
\left[\mathrm{Re}z_1 \, x_\infty-(\mathrm{Re}z_1)^2-(\mathrm{Im}z_1)^2\right]
/(x_\infty-\mathrm{Re}z_1)$
and
$t_1=(\mathrm{Im}z_1)^2x_\infty^4/\{4
[(\mathrm{Re}z_1-x_\infty)^2+(\mathrm{Im}z_1)^2]^2\}$.
The map $G_{t_1,\,\xi_0}$ is then applied to the points resulting
in a new sequence, by one element shorter: $z'_{k} =
G_{t_1,\,\xi_0}(z_{k+1})$ with $k=1\ldots N$.
The operation is iterated on the new
subsequence of points until one obtains the full set of $t_{k}$
and $\xi_{k}$ that gives a piecewise constant approximation of the
driving function.

The result of this procedure is an ensemble of $\xi(t)$ whose
statistics converge, for $l_f^2 \!\lesssim \!\kappa t \!\lesssim
\!l_\tau^2 $ to a Gaussian process with the variance
$\langle \xi^2(t) \rangle\! =\! \kappa t$ and $\kappa\!=\!4\!\pm
0.2$, as shown in Figure~\ref{fig:4}. We conclude that, within
statistical errors,
 zero-temperature isolines in the inverse cascade of
SQG turbulence are locally SLE$_4$ curves. This applies also to
other $T$ contours provided that $T \ll T_{rms}$. Moreover, in the
limit of very large system size where $T({\bm r})$ tends to a
self-affine field and $T_{rms}$ diverges, we expect all iso-level
loops to be statistically equivalent (another application of SLE
to non-equilibrium systems has been found recently for spin
glasses \cite{Hast}).

Remark that Surface Quasi-Geostrophic  and Navier-Stokes systems
belong to a class that is uniquely specified by the transport
equation~(\ref{eq:1}) and by a linear, scale-invariant, local in
time  relationship between the advected field
 and the streamfunction $\psi({\bm x},t)= \int d{\bm y}\,
 |{\bm x}-{\bm y}|^{\alpha-2} T({\bm y},t)$ \cite{PHS94}.
 SQG dynamics corresponds to
$\alpha=1$, Navier-Stokes equation to $\alpha \to 2$, $T$ being
vorticity. The large-scale limit of the Charney-Hasegawa-Mima
model \cite{CHM}, relevant  for atmospheric and plasma turbulence,
corresponds to $\alpha=-2$. Dimensional arguments for the inverse
cascade  in this class of models give $\delta_r T\propto r^{H}$ with
$H=(2-2\alpha)/3$. This exponent can be used to infer the
dimension of the contour loops $(3-H)/2=(7+2\alpha)/6$ \cite{KH95}
(for $0<H<1$ and $-1/2 <\alpha<1$) and thus conjecture that they
converge to SLE$_\kappa$ curves with $\kappa=4(1+2\alpha)/3$.

Let us stress that the temperature field in our model has
non-Gaussian statistics, see \cite{CCMV04} and Fig.~\ref{fig:0}. A
similar non-Gaussian form with logarithmic moments holds for the
height function built on independently oriented loops from the
$O(n)$ model \cite{CZ,C06}
--- yet it requires $n\not=2$ and $\kappa\not=4$ (for $\kappa=4$
the statistics is Gaussian). Should our field belong to this
class, the difference $\kappa-4$ would be much larger than our
$5\%$ margin of error (as can be inferred comparing
Fig.~\ref{fig:4} with the results of \cite{C06}). It remains to be
understood how such a non-Gaussian field can have isolines with
the same statistics as the isolines of the Gaussian free field.

\begin{figure}[!hb]
\includegraphics[angle=270,width=8.5cm]{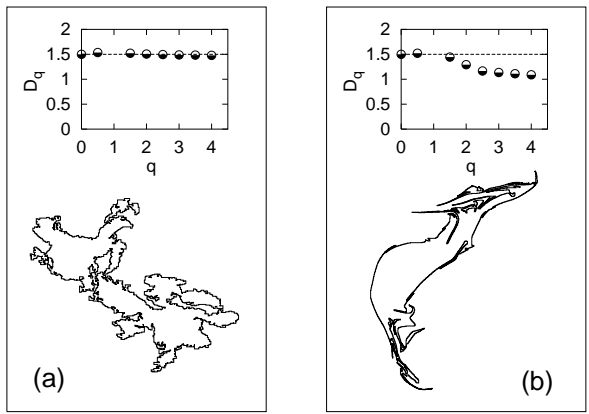}
\caption{{\it (a)\/}  A typical loop in SQG inverse cascade.
Inset: generalized fractal dimensions $D_q=(q-1)^{-1} \lim_{r\to \infty}
\ln Z_q(r)/\ln r$ where $Z_q(r)=\sum \mu_i(r)^q$ and the
sum runs over a set of $N(r)$ non-overlapping $r$-boxes covering
the curve, and $\mu_i(r)$ is the mass of the curve inside the
$i$-th box. {\it (b)\/}  A typical vorticity loop in the direct
cascade of 2d Navier-Stokes turbulence, obtained from a
pseudo-spectral simulation with 1024$^2$ lattice points. The
spectrum of generalized dimensions shows a clear dependence on $q$
(multifractality). It decreases from the fractal dimension
$D_0\approx 3/2$ to reach values $D_q \approx 1$ for large $q$, as
a result of the existence of long, almost one-dimensional segments
of the loops.} \label{fig:5}
\end{figure}

Let us briefly compare our findings with other turbulent systems.
In the direct cascade of 2d Navier-Stokes turbulence, the
vorticity field has logarithmic correlations  ($H=0$) and is
characterized by very weak, if any, deviations from
self-similarity; vorticity isolines have dimension $D=3/2$, just
as in the inverse cascade of SQG turbulence. However, the
similarities end here. The loops in the direct cascade,
shown in Figure~\ref{fig:5}, are not SLE  curves since they are
not even scale invariant as seen from the multifractal spectrum in
the inset in Fig.~\ref{fig:5}(b). Therefore, it appears that
inverse cascades are more akin to statistical mechanics systems
and more appropriate for conformal invariance. Another necessary
condition may be strong nonlinearity since the turbulence of
weakly interacting waves is generally not conformal invariant
(except when it has logarithmic correlations in 2d). Another
relevant example is a passive scalar in a spatially smooth random
flow (Batchelor regime), which also has logarithmic correlation
functions \cite{Bat}. In this case, cascade direction depends on
the compressibility of the flow \cite{FGV}. By a straightforward
applications of the formulas from \cite{BCKL,CKV} one can show
that the four-point correlations in the Kraichnan model are not
conformal invariant  in either direct or inverse cascade.

To conclude, we have found the second example of conformal
invariance in turbulence thus showing that neither Navier-Stokes
turbulence nor percolation are unique in this respect. Both cases
correspond to inverse cascades. In the direct/inverse cascade  we
study statistics at the scales which are respectively
smaller/larger than the pumping correlation scale. It is thus not
surprising that the direct cascade is sensitive to the
statistics of the pumping \cite{FS,FGV};  even when there is scale
invariance, conformal invariance is absent as shown here. In
inverse cascades, short correlated random force imposes some
degree of locality and yet conformal invariance  is a remarkable
example of emerging symmetry since our systems are dynamically
nonlocal and far from equilibrium.
It remains an open question, here as in the
Navier-Stokes case, whether conformal invariance extends to some
field correlation functions, and how to identify candidate primary
fields upon which a conformal field theory for the inverse cascade
can be built.

This work has been partially supported by ISF and ANR BLAN06-3-134460,
and by CNISM for computational resources.

\end{document}